# Efficient Algorithmic Techniques for Several Multidimensional Geometric Data Management and Analysis Problems

Mugurel Ionuţ Andreica
Politehnica University of Bucharest, Romania, mugurel.andreica@cs.pub.ro

**Abstract:** *In this paper I present several novel, efficient, algorithmic techniques for solving some multidimensional geometric data management and analysis problems. The techniques are based on several data structures from computational geometry (e.g. segment tree and range tree) and on the well-known sweep-line method.*

**1. Introduction**

Managing, analyzing and extracting information from multidimensional data are very important activities in a broad range of domains, like business planning and integration, scientific research, statistics and even distributed systems. In this paper I consider problems like the largest empty circle and hyper-rectangle in a given set of points, deciding containment in sets of circles and hyper-rectangles and computing maximum weight subsequences of points, which must be accepted by a modified non-deterministic finite automaton. Sections 2, 3 and 4 present the results regarding these problems and in Section 5 I discuss related work and I conclude.

**2. Largest Empty Circle and Hyper-Rectangle with Fixed Aspect Ratio**

We are given $n$ points in the plane, at coordinates $(x_i, y_i)$. We want to find a circle $C_{opt}$ of maximum radius, completely located inside another circle $C(xc, yc, rc)$, s.t. $C_{opt}$ contains none of the $n$ points inside of it. The solution is based on binary searching the maximum radius $R$ of $C_{opt}$. Once the radius $R$ is fixed in the binary search, we need to solve a decision problem. We construct a circle of radius $R$ around each of the $n$ points and „shrink" the circle $C$ to a circle $C'(xc, yc, rc-R)$. We need to decide if the $n$ circles cover completely the circle $C'$. If they do, then we need to test a smaller value of $R$; otherwise, we will test a larger value. We will compute the intersections between every two of the $n$ circles (in $O(n^2)$ time). We then construct the set of x-coordinates $S$, consisting of the x-coordinates of the intersection points and the points $\{x_i - R,\ x_i,\ x_i + R\}$ ($1 \leq i \leq n$). We sort the points in $S$: $xs_1 < xs_2 < ... < xs_M$ ($M = O(n^2)$ is the total number of distinct x-coordinates in $S$). The points define $M-1$ vertical slabs (slab $i$ is between $xs_i$ and $xs_{i+1}$). We compute the intersection of each vertical slab with every circle. If a circle intersects the slab, then it defines a kind of trapezoid with „rounded" contours at the top and the bottom. Because all the intersections between circles occur on the slab boundaries, the top and bottom contours of the trapezoids do not intersect. Thus, we can reduce the intersection of every circle with the vertical slab $i$ to the intersection with the vertical line $x=(xs_i+xs_{i+1})/2$. We obtain a set of at most $n+1$ intervals on this line. If *C'* intersects the line on an interval *I'*, we need to verify if the intervals corresponding to the *n* circles completely cover *I'*. We can do this by sorting the endpoints of the intervals ($yint_1<yint_2<...<yint_k$) and then traversing the endpoints. During the traversal we maintain a counter, denoting the number of open intervals covering each „primitive" interval *[$yint_j$, $yint_{j+1}$]* (we increment the counter by *1* at every left endpoint and decrement it by *1* at every right endpoint). If some primitive interval included in *I'* is covered by only one interval, then *I'* (*C'*) is not completely covered by the other intervals (*n* circles). The time complexity of this algorithm is $O(n^3 \cdot log(n))$. This technique can also be used for computing the area of the union of *n* circles. Within each vertical slab, we compute the union of the intervals on the line $x=(xs_i+xs_{i+1})/2$. For each interval in the union, we store the identifiers of the circles to which its bottom and top endpoints belong. Afterwards, we extend the intervals in the union to trapezoids in the vertical slab. Each trapezoid can be decomposed into a rectangle and two „rounded" top and bottom parts. The area of the rectangle is easy to compute; the area of a „rounded" part is computed by knowing to which circle *q* it belongs and the angle interval it spans on *q* (with this information, its area is the difference between the area of a circle sector and two triangles). Of course, we can also compute the union of polygonal objects this way. We compute the set *S* as the union of the x-coordinates of all the intersection points between every two polygonal objects and the x-coordinates of the polygons' vertices. The intersection between a polygon and a vertical slab is a set of disjoint trapezoids. In order to avoid complex intersection cases, we can replace every polygon by a collection of disjoint triangles covering the entire polygon (triangulation). Since no intersections occur inside a slab, we can consider only the intersections with a vertical line in the middle of the slab, compute the union of the intervals and for each resulting interval in the union, store the polygon edge to which its two endpoints belong. We then extend each interval in the union to the entire slab, obtaining a trapezoid, whose area is easy to compute. The time complexity of this approach is $O(M \cdot n \cdot log(n))$, where *M* is the total number of distinct x-coordinates in the set *S* and *n* is the total number of edges of the polygons. For some special cases of polygons, *M* may be much less than $O(n^2)$. For instance, for axis-aligned rectangles and several other types of polygons, $M=O(n)$. An improvement for this technique (both for circle and polygons) consists of not re-sorting the *curves* (polygon edges and circle contours) within every vertical slab, because their order changes only slightly. For instance, if at most two *curves* intersect at a given point, then, when moving from slab *i* to slab *i+1*, we only need to swap the order of two adjacent curves, insert a new curve or remove an old one. If we maintain the curves intersecting the current slab in a balanced tree, we can perform every operation in $O(log(n))$ time. Thus, the time complexity becomes

$O(M \cdot (n+log(n)))$ (because we can traverse the curves in the tree in $O(log(n)+n)$ time and, for each curve, compute its intersection with the slab).

We now want to find a largest volume d-dimensional hyper-rectangle $HR$, fully included inside a hyper-rectangle $R(xa_1, xa_2, ..., xa_d, xb_1, xb_2, ..., xb_d)$ which contains none of the $n$ given points (a point $i$ is a tuple $(x_{i,1}, x_{i,2}, ..., x_{i,d})$). $R$ contains all the points $(xp_1, ..., xp_d)$, s.t. $xa_j \leq xp_j \leq xb_j$ ($1 \leq j \leq d$). Let's assume that $l_1, ..., l_d$ are the lengths of $HR$ in each of the $d$ dimensions. We must also have $l_j = l_1 \cdot f_j$ ($1 \leq j \leq d$) (the *fixed aspect ratio* property), with $f_1 = 1$ and $f_j$ ($2 \leq j \leq d$) arbitrary non-negative real numbers. We will binary search the length $l_1$ of $HR$. We now need to perform a feasibility test for the length $l_1$. For each point $i$ ($1 \leq i \leq n$) we build a hyper-rectangle $HP_i = (xp_1 - l_1 \cdot f_1, xp_2 - l_1 \cdot f_2, ..., xp_d - l_1 \cdot f_d, xp_1, xp_2, ..., xp_d)$. We also "shrink" the hyper-rectangle $R$ to $R'(xa_1, xa_2, ..., xa_d, xb_1 - l_1 \cdot f_1, xb_2 - l_1 \cdot f_2, ..., xb_d - l_1 \cdot f_d)$. If the union of the rectangles $HP_i$ covers $R'$, then $l_1$ is too large and we need to consider a smaller value; otherwise, we will consider a larger value. We will clip the hyper-rectangles $HP_i$, s.t. they are located inside $R'$, i.e. for each dimension $j$ ($1 \leq j \leq d$), the left side of $HP_i$ is at $max\{xp_j - l_1 \cdot f_j, xa_j\}$ and the right side at $min\{xp_j, xb_j - l_1 \cdot f_j\}$. We can now compute the volume $V$ of the union of the $n$ hyper-rectangles. If $V$ is equal to the volume of $R'$, then $R'$ is completely covered by the $n$ hyper-rectangles; otherwise, it is not. We can compute the volume of $n$ axis-aligned hyper-rectangles in $O(n^{d-1} \cdot log(n))$ in the following way. We sort the $2 \cdot n$ endpoints in each dimension. We traverse the "primitive" intervals between the endpoints of the $d^{th}$ dimension and maintain a set with the hyper-rectangles which completely cover each primitive interval (a hyper-rectangle is added to the set when its left endpoint is encountered and removed from the set when its right endpoint is encountered). For every primitive interval, we will compute the (d-1)-dimensional volume of the hyper-rectangles completely covering it, multiply it to the length of the primitive interval and add it to the total volume. If we use this approach all the way to $d=1$, the time complexity is $O(n^d + d \cdot n \cdot log(n))$; however, for $d=2$, there exists an $O(n \cdot log(n))$ algorithm [1].

## 3. Circle and Hyper-Rectangle Containment Problems

Let's consider that we are given $n$ circles, by the coordinates of their centers $(x_i, y_i)$ and their radiuses $r_i$, s. t. every two circles are either totally disjoint or one of them is included inside the other. We want to decide, for each circle, if it contains (is contained by) another circle. We will solve this problem by a sweep-line algorithm. First, we divide each circle $i$ into two half-circles, $HCL_i$ and $HCR_i$ (left and right). If circle $i$ is included inside circle $j$, then at least one of the following holds: $HCL_i$ is included in $HCL_j$ or $HCR_i$ is included in $HCR_j$. Thus, we can restrict our attention to half-circles and run the sweep-line algorithm twice. I will describe an $O(n \cdot log(n))$ algorithm for sweeping the right half-circles from left to right (symmetrically, we will sweep the left half-circles from right to left). We will maintain a balanced tree

(e.g. AVL tree) with the disjoint intervals representing the intersections of the half-circles with the current position of the vertical sweep-line. Since the intervals are disjoint, the tree behaves as if numerical values equal to the intervals' left endpoints are stored in it (the right endpoints do not make any difference). We consider the set of x-coordinates given by the union of the sets $\{x_i, x_i+r_i\}$, sorted from left to right (including duplicates). Whenever we reach a new x-coordinate, we need to update (some of) the intervals in the balanced tree. For each coordinate we maintain information about the corresponding half-circle $i$, i.e. if it „*begins*" at this coordinate ($x_i$) or if it „*ends*" there ($x_i+r_i$). When a half-circle $i$ *begins*, it covers the interval $I_i=[y_i-r_i, y_i+r_i]$ on the sweep-line. We traverse all the intervals in the tree which intersect $I_i$. We can find the first such interval $I_{first}$ in $O(log(n))$ time (it is either the one with the largest left endpoint $le \leq y_i-r_i$ or the one with the smallest left endpoint $le > y_i-r_i$). Then, we iteratively find the successor of every interval, starting from $I_{first}$ (in $O(log(n))$ time). Each such interval $I_j'$ is the intersection of a half-circle $j$ with a previous position of the sweep-line. We compute the new intersection $I_j$ of $j$ with the current position of the sweep-line, remove $I_j'$ from the tree and insert $I_j$. If $I_j$ ($I_i$) is included in $I_i$ ($I_j$), then circle $i$ ($j$) includes circle $j$ ($i$) and we remove $I_j$ (nothing) from the tree. If $I_i$ is not included in any $I_j$, we insert $I_i$ in the tree. When a half-circle *ends*, we remove its interval from the tree (if the tree contains such an interval).

We now consider the same problem for the case of $n$ d-dimensional hyper-rectangles (defined like in the previous section) which may intersect partially. We will sort the $2 \cdot n$ endpoints along dimension $d$ and sweep the rectangles with a (d-1)-dimensional hyperplane. When we encounter the left endpoint of a hyper-rectangle $i$, we will consider the (2·d-2)-dimensional point $p_i=(xa_{i,1}, ..., xa_{i,d-1}, xa_{i,d}, xb_{i,1}, ..., xb_{i,d-2})$ with weight $wl_i=xb_{i,d-1}$ and *activate* it into a (2·d-2)-dimensional range tree *RT*. A k-dimensional range tree [3] is a balanced binary tree, where each leaf stores a *k-D* point; the points are sorted according to the $k^{th}$ coordinate from the leftmost leaf to the rightmost one. Each internal node stores all the points contained in the leaves of its subtree; thus, every point is stored in $O(log(n))$ tree nodes. Each tree node $q$ of the range tree contains a (k-1)-dimensional range tree $T_q$, storing all the points corresponding to node $q$. For $k=2$, each tree node $q$ maintains a segment tree $T_q$ instead of a 1D range tree. The segment tree is built upon the $1^{st}$ coordinate of the points and each node of $T_q$ maintains the largest weight *maxw* of a point contained in its subtree (points are contained only at leaf nodes). We will use the segment tree as in [2], for range maximum queries and point updates.

The range tree *RT* is built over all the *n* points $p_i$, with initial weights equal to -∞. When the left endpoint (in dimension $d$) of a hyper-rectangle is encountered, we modify the weight of $p_i$ to $wl_i$; when we encounter a right-endpoint, we modify $p_i$'s weight back to -∞. In order to change the weight of a point $p_i$, we consider all the $O(log(n))$ nodes $q$ of RT containing $p_i$ and call the procedure for the range trees $T_q$ of these nodes. When we arrive at a node $q$ in a 2D range tree, we update the weight associated to the $1^{st}$ coordinate of the point in the segment tree $T_q$ (as in [2]). *RT* provides a function *findMaxW((xa$_1$,xb$_1$), ..., (xa$_{2 \cdot d-2}$, xb$_{2 \cdot d-2}$))*, which returns the maximum weight of a point contained in *RT* and the given ranges. The search in a *k-D* range tree proceeds as follows. It finds $O(log(n))$ nodes (a canonical subset) containing all the points whose $k^{th}$ coordinate is contained in the $k^{th}$ range and calls the function for the *(k-1)-D* range trees $T_q$ contained in these nodes. When $k=2$, we just call the range query function of the segment tree $T_q$, obtaining the maximum weight in the corresponding range (it is preferable to normalize the range *(xa$_1$,xb$_1$)* to a range of indices *[a,b]*, $1 \le a \le b \le u$, where *u* is the number of leaves in $T_q$; we can normalize it by binary searching the index *a* of the smallest coordinate larger than $xa_1$ and the index *b* of the largest coordinate smaller than $xb_1$ in $T_q$).

Whenever we encounter the right endpoint (in dimension *d*) of a hyper-rectangle *i* during the sweep, we call *RT.findMaxW((-∞,xa$_{i,1}$), ..., (-∞,xa$_{i,d-1}$), (-∞,xa$_{i,d}$), (xb$_{i,1}$,+∞), ..., (xb$_{i,d-2}$,+∞))*, obtaining the value *mw*. If *mw<xb$_{i,d-1}$*, then hyper-rectangle *i* is not included in any hyper-rectangle. Otherwise, by easily modifying the described function to return the hyper-rectangle *j* whose weight $wl_j$ is maximum in the given range, we decide that hyper-rectangle *j* contains hyper-rectangle *i*.

The time complexity of this algorithm is $O(n \cdot log^{2 \cdot d-2}(n) + n \cdot log(n))$, for any dimension $d \ge 1$. We should note that an $O(n \cdot log^{2 \cdot d-1}(n))$ solution exists, too. We assign to every hyper-rectangle a (2·d)-dimensional point $r_i=(xa_{i,1}, ..., xa_{i,d}, xb_{i,1}, ..., xb_{i,d})$ and construct a range tree over these points (we do not assign weights to the points). Then, for each rectangle, we range count the number of points in the range *(-∞,xa$_{i,1}$)* x ... x *(-∞,xa$_{i,d}$)* x *(xb$_{i,1}$,+∞)* x ... x *(xb$_{i,d}$,+∞)* in $O(log^{2 \cdot d-1}(n))$ time. In order to decide if a hyper-rectangle *i* contains another hyper-rectangle, we range count the number of points in the range *(xa$_{i,1}$,+∞)* x ... x *(xa$_{i,d}$,+∞)* x *(-∞,xb$_{i,1}$)* x ... x *(-∞,xb$_{i,d}$)*.

## 4. Maximum Weight Subsequence Accepted by a NFA

We are given a sequence of *n* d-dimensional points *pt(i)=(x(i,1), ..., x(i,d))* ($1 \le i \le n$), each point having a weight *w(i)*. We are also given a special kind of non-deterministic finite automaton (NFA) with *m* states *q(0), ..., q(m-1)* (each state can be an *initial* state, a *final* one, both or none) and a set *E* of directed edges. An edge $e_k=(q(i),q(j))$ is labeled with *d* intervals *[a(k,1), b(k,1)], ..., [a(k,d), b(k,d)]*. We want to find a subsequence of points *s(0)<s(1)<...<s(l)* (not necessarily contiguous), composed of any number of points, such that: *(i)* there exists a sequence of states *q(u(0)), ..., q(u(l))*, with the property that we can „pass" from each state *q(u(i-1))* to *q(u(i))* using the point *s(i)*, *q(u(0))* is an *initial* state and *q(l)* is a *final* state; *(ii)* the sum of the weights of the points in the sequence is maximum. We can „pass" from a state *q(u(i-1))* to *q(u(i))* ($1 \le i \le l$), if an edge $e_k=(q(u(i-1)),q(u(i))) \in E$ and $a(k,p) \le x(s(i),p)-x(s(i-1),p) \le b(k,p)$. We will compute a table $W_{max}(i,j)$=the maximum weight of a subsequence ending at point *i* and reaching the state *q(j)*. Then, we will compute $max\{W_{max}(i,j)|1 \leq i \leq n, 0 \leq j \leq m-1, q(j)$ is a final state$\}$. Computing the actual sequence can be performed by tracing back the way the $W_{max}(i,j)$ values were computed. We define $W_{max}'(i,j)=w(i)+max\{W_{max}(i',j')| 1 \leq i'<i, 0 \leq j' \leq m-1, \exists\ e_k=(q(j'),q(j)) \in E$ s.t. $a(k,p) \leq x(i,p)-x(i',p) \leq b(k,p)$ for $1 \leq p \leq d\}$. We have $W_{max}(i,j)=max\{w(i), W_{max}'(i,j)\}$ if *q(j)* is an initial state and $W_{max}(i,j)=W_{max}'(i,j)$, otherwise. In order to compute the values $W_{max}$ and $W_{max}'$, we will traverse the points from *1* to *n* and maintain *m* d-dimensional range trees *RT(k)*, $0 \leq k \leq m-1$. Each range tree is constructed over the whole set of points *pt(i)*, with initial weights equal to -∞. After computing the values $W_{max}(i,j)$, we change the weight of *pt(i)* in *RT(j)*, to $W_{max}(i,j)$. In order to compute $W_{max}'(i,j)$, we consider every state *j'* s.t. $e_k=(q(j'),q(j)) \in E$ and find the maximum weight of a point in *R(j')*, in the range *[x(i,p)-b(k,p), x(i,p)-a(k,p)]* $(1 \leq p \leq d)$. The range tree we use was described in the previous section. If *d=1*, the range tree becomes a segment tree (also mentioned previously). The time complexity of this approach is $O(n \cdot (m+|E|) \cdot log^d(n))$. As an application, consider: a NFA with *1* state *q(0)* (both *initial* and *final*) and one edge $e_1=(q(0),q(0))$, labeled *[ε,∞]*; *d=1* and *w(i)=1*. The algorithm computes the *longest increasing subsequence* in $O(n \cdot log(n))$ time, relative to the *x(i,1)* values (where *0<ε<<1* is an arbitrarily small constant); with *2* states (both *initial* and *final*) and two edges $e_1=(q(0),q(1))$ and $e_2=(q(1),q(0))$, labeled *[ε,∞]* and *[-∞,-ε]*, we can compute the *longest alternating subsequence*.

## 5. Conclusions & Related Work

Computing largest empty objects, deciding containment for sets of hyper-spheres or hyper-rectangles and computing optimal subsequences are fundamental problems in computational geometry [1,3,4,5], with applications to several multidimensional data analysis problems. My original motivation behind this work consists of the analysis of data gathered from distributed systems with nodes whose identifiers are mapped into a d-dimensional metric space. In this case, the size of a largest empty circle or hyper-rectangle offers hints about the distribution of the identifiers in the metric space. Circle and hyper-rectangle containment problems are related to the case when the nodes' identifiers are not just points, but also circles or rectangles, whose area is proportional to the bandwidth or computational power of the node.